\documentclass[aps,prd,notitlepage,nofootinbib,nobibnotes,superscriptaddress,amsmath,amssymb,preprintnumbers]{revtex4-1}
\usepackage{graphicx} 
\usepackage{amsmath}
\usepackage{dcolumn}
\usepackage[colorlinks=true,linktocpage=true,%
  linkcolor=blue,citecolor=blue,urlcolor=blue]{hyperref}
\usepackage{color}
\usepackage{slashed}
\usepackage     [utf8]                  {inputenc}

\newcommand{\be}{\begin{equation}}
\newcommand{\ee}{\end{equation}}

\definecolor{oscar}{RGB}{22, 156, 172}

\definecolor{vermilion}{rgb}{0.89, 0.26, 0.2}

\newcommand{\rmd}{\mathrm{d}}
\newcommand{\rme}{\mathrm{e}}
\newcommand{\rmi}{\mathrm{i}}

\newcommand{\calN}{\mathcal{N}}

\newcommand{\bperp}{\boldsymbol{b}_\perp} % transverse b
\newcommand{\xp}{\boldsymbol{x}_\perp} % transverse x
\newcommand{\kgp}{\boldsymbol{k}_{\gamma\perp}} % k photon perp
\newcommand{\kp}{\boldsymbol{k}_\perp}

\newcommand{\qp}{\boldsymbol{q}_\perp}

\newcommand{\Php}{\boldsymbol{P}_{h\perp}}

\begin{document}
\date{\today}
\preprint{ZTF-EP-22-01}

\title{Isolated photon-hadron production in high energy $pp$ and $pA$ collisions at RHIC and LHC}
\author{Sanjin Beni\' c}
\affiliation{Department of Physics, Faculty of Science, University of Zagreb, Bijenička c. 32, 10000 Zagreb, Croatia}
\author{Oscar Garcia-Montero}
\affiliation{Institut f\" ur Theoretische Physik, Goethe Universit\" at,
Max-von-Laue-Strasse 1, 60438 Frankfurt am Main, Germany}
\author{Anton Perkov}
\affiliation{Department of Physics, Faculty of Science, University of Zagreb, Bijenička c. 32, 10000 Zagreb, Croatia}
              
\begin{abstract} 
We compute the isolated photon production in association with a charged hadron at mid rapidity in $pp$ and $pA$ based on the Color Glass Condensate (CGC) framework of high energy QCD where, for the first time, we incorporate the Sudakov effect of soft gluon emissions. Our results are based on the leading order $q g \to q \gamma$ channel in the CGC framework and confronted with the recent data from RHIC and LHC concerning the angular distributions and out-of-plane transverse momentum distributions. We find that, while the CGC computation alone results in too narrow distributions, with the help of the Sudakov effect, we can get a satisfactory description of the data. With this as a benchmark, we provide predictions for the magnitude of the nuclear effect brought by the phenomena of gluon saturation in the CGC. 
\end{abstract}

\maketitle

\section{Introduction}\label{sec:intro}

Photon-hadron production in high energy $pp$ and $pA$ collisions has been put forward \cite{Jalilian-Marian:2005qbq,Jalilian-Marian:2008jjn} as a promising probe of the small-$x$ hadron wavefunction \cite{Iancu:2003xm,Jalilian-Marian:2005ccm,Gelis:2010nm}. The distinguished character of this process lies in the fact that a well isolated photon does not participate in strong interactions and therefore the uncertainties related to hadronization of the final state are reduced compared to the more abundant di-hadron ($hh$) production.
The isolated photon component is defined to be separated by a suitable isolation algorithm from the so-called \emph{decay photons}, which are radiated from the $\pi^0 \to \gamma\gamma$ decays. This, in turn, suppresses photons fragmenting off jets in the course of hadronization. Recently, the PHENIX collaboration at RHIC \cite{PHENIX:2016zxb,PHENIX:2018trr} and the ALICE collaboration at the LHC \cite{Acharya:2020sxs} reported on the measurements of the cross section of an isolated photon in association with an unidentified charged hadron in $pp$ and $pA$ that is within the kinematic region potentially sensitive to small-$x$ physics. In this work we are motivated to pursue the phenomenological implications of those measurements for the first time.

A long standing hypothesis in high energy QCD is that the rapid growth of gluon radiation inside the hadron wavefunction at small-$x$ is balanced out by a gluon recombination process, which leads to the phenomena of gluon saturation. The theoretical framework behind this basic picture is called the Color Glass Condensate (CGC) \cite{Iancu:2003xm,Jalilian-Marian:2005ccm,Gelis:2010nm} where the energy, or small-$x$, enhanced logarithms $\sim \alpha_S \log (s) \sim \alpha_S\log \left(1/x\right)$, can be consistently resummed at each order in the strong coupling $\alpha_S$. The resulting gluon radiation in CGC becomes concentrated around a dynamically generated transverse momentum - the saturation scale $Q_S$, that is build up from multiple scattering of partons on the dense target.  In the context of the simultaneous production of two particles, the back-to-back transverse momenta kinematics of the conventional leading order $2\to 2$ partonic process gets disrupted in the CGC. This becomes more dramatic for a heavy nuclei, where a scaling $Q_S^2 \sim A^{1/3}$ is expected.

It is with the above picture in mind that $\gamma h$ \cite{Jalilian-Marian:2012wwi,Rezaeian:2012wa,Rezaeian:2016szi,Benic:2017znu,Goncalves:2020tvh} and $\gamma^* h$ \cite{Stasto:2012ru,Basso:2015pba,Basso:2016ulb} production in $pA$ (and recently also the related $\gamma-{\rm jet}$ production in $eA$ \cite{Kolbe:2020tlq}) have been explored as possible pathways to a phenomenological validation of the CGC. Saturation effects become prominent when the imbalance momenta of the photon ($\kgp$) and the underlying final state parton ($\qp$) is of the order of the saturation scale $k_\perp \equiv |\kgp + \qp| \sim Q_S$. In terms of the distribution over the azimuthal angle $\Delta\phi \equiv \phi_\gamma - \phi_h$, the general expected feature is a broadening of the away side peak, $\Delta \phi = \pi$, as the kinematics condition $k_\perp \to Q_S$ becomes satisfied.

While both $\gamma h$ and $hh$ production are understandable with such a simple physics argument of the broadening of the away side peak, further investigation into the two processes revealed that they are sensitive to two different unintegrated gluon distributions (UGD)s \cite{Kharzeev:2003wz}. At leading order, $\gamma h$ production is probing the gluon dipole distribution, while $hh$ production is instead probing a combination of the dipole and the Weisz\" acker-Williams distribution \cite{Dominguez:2010xd,Dominguez:2011wm}. These two UGDs have a completely different behavior at small $k_\perp$ (see e.~g. \cite{Dominguez:2011wm} and also Eq.~\eqref{eq:dpww} in the following section). Thus, according to CGC, the theoretical prediction for $\gamma h$ production at the away side peak may look nothing like the prediction for $hh$ production. Indeed this is another reason to explore $\gamma h$ production in CGC. 

Alternatively, the broadening of the away side peak may %have nothing to do with 
be completely unrelated to the initial state, but simply be induced by the soft gluon radiation, the so-called Sudakov effect \cite{Sudakov:1954sw,Dokshitzer:1978hw,Parisi:1979se,Collins:1984kg,Collins:2011zzd}. Soft gluon radiation generates double logs as $\alpha_S\log^2(k_\perp/Q)$, where $Q$ is a hard scale in the process, and therefore becomes enhanced when the underlying individual partonic transverse momenta is hard, $k_\perp \ll Q$. For the high energy kinematics case when in addition $k_\perp \sim Q_S$ holds, saturation effects and Sudakov effects become two contributing mechanisms and so clearly both should be taken into account. A recent development \cite{Mueller:2012uf,Mueller:2013wwa} showed that it is possible to perform a simultaneous resummation of both the Sudakov logs and the small-$x$ logs in a consistent way. On the other hand, since $Q_S^2 \sim A^{1/3}$ for a heavy nuclei, it is expected the saturation effect would become more important in the $pA$ collision, while both saturation and Sudakov effect are needed for a proper description in $pp$ collisions. This is indeed confirmed by recent studies in $hh$ production \cite{Stasto:2018rci,vanHameren:2020rqt} but also in dijet \cite{vanHameren:2019ysa}, $Z$-boson \cite{Marquet:2019ltn} and $Z$-jet \cite{vanHameren:2015uia} productions as well as in related processes in $eA$ collisions \cite{Zheng:2014vka,Zhao:2021kae,vanHameren:2021sqc}, which prompts us to take into account both the small-$x$ and Sudakov effects in order to establish a realistic baseline for a proper extraction of the appropriate nuclear modifications based on a comparison of the obtained results in $pp$ and $pA$ collisions.

The main purpose of this work is to perform a numerical computation of the isolated photon-hadron cross section within the CGC framework, while taking into account Sudakov effects for the first time. In Sec.~\ref{sec:framework} we explain our approach with the main formulas given in \eqref{eq:W} and \eqref{eq:incgammahlo2}. Our numerical results, shown in Sec.~\ref{sec:results}, concern the azimuthal angle ($\Delta \phi$) distributions measured at RHIC at 200 GeV and 510 GeV in $pp$ and the LHC at 5.02 TeV in $pp$ and $p$Pb and also the out-of-plane transverse momentum distributions measured at RHIC. Our main finding is that while the CGC results alone produce too narrow distributions compared to the data a reasonable agreement with the data is possible when the Sudakov effect is taken into account. Additionally we provide predictions for the magnitude of the nuclear effect in $pA$ vs $pp$ also arguing in favour of more symmetric transverse momentum kinematics of the $\gamma h$ system where the effect of gluon saturation would be better resolved. In the final Section~\ref{sec:conc} we summarize our findings while also providing an outlook to further theoretical investigations.

\section{Theoretical framework}
\label{sec:framework}

In this section we give a quick recap of the main formulas that will be used in obtaining our numerical results for $p (P_p) A (P_A) \to h^{\pm} (P_h)\gamma (k_\gamma) X$ production. Throughout this work, we will use  $p$ for the projectile proton, $A$ for the target nucleus (or proton in $pp$ collisions). We will also use the shorthand notation $h^\pm \equiv h^+ + h^-$ to express observables which are obtained by adding the contributions of positive and negative unidentified charged hadrons. We will be using the leading order formulas for $\gamma h$ production - at mid-rapidity the dominant channel is $q(p)g(k) \to q(q) \gamma(k_\gamma)$, which we firstly address below in the CGC framework, and subsequently incorporate soft gluon resummation in the following subsection. The final formula that we will be using in our numerical computations is given in \eqref{eq:W} and \eqref{eq:incgammahlo2}.

\subsection{Photon-hadron cross section in the CGC framework}\label{sec:phXsec}

The following CGC formulas are computed using a dilute-dense framework \cite{Iancu:2003xm,Jalilian-Marian:2005ccm,Gelis:2010nm} where a dilute projectile parton passes through a dense (nuclear) target described by a classical gluon field. In the dilute-dense framework an all-order scattering on the target is taken into account building up a finite transverse momentum in the final state, while the dilute projectile is treated order-by-order in perturbation theory.
The leading order $pA\to h\gamma X$ inclusive CGC cross section %that 
can be straightforwardly obtained from the underlying partonic $q g \to q\gamma$ \cite{Kopeliovich:1998nw,Gelis:2002ki,Baier:2004tj} channel. %cross section. 
In the massless quark limit it simplifies %, $m_q \rightarrow 0$, 
to the following expression, 
\be
\frac{\rmd \sigma}{\rmd^2 \kgp \rmd \eta_\gamma\rmd^2 \Php\rmd \eta_h} = (\pi R_A^2) \sum_q\int_0^1 \frac{\rmd z_h}{z_h^2} D_q(z_h,\mu^2)\frac{e_q^2 N_c}{8\pi^4} x_p f_q(x_p,\mu^2)\kp^2\tilde{\calN}_{A,Y_A}(\kp)\hat{\sigma}\,,
\label{eq:incgammahlo}
\ee
where $f_q(x_p,\mu^2)$ is the collinear quark distribution function, $D_q(z_h,\mu^2)$ is the relevant collinear fragmentation function of a quark with flavor $q$ and momenta $q^\mu = P_h^\mu/z_h$ to a particular hadron species at a factorization scale $\mu^2$. In this work we are using the CTEQ6M quark distributions \cite{Pumplin:2002vw} and the DSS fragmentation functions \cite{deFlorian:2007aj}.

We have $\kp \equiv \kgp + \qp$ as the imbalance momentum. Furthermore, $(\pi R_A^2)$ is the target area. The hard factor $\hat\sigma$ is given as
\be
\hat{\sigma} = \frac{\alpha_e}{2 N_c}\frac{P_{q\gamma}(z)}{q\cdot k_\gamma} \frac{z^2}{\kgp^2}\quad \text{with} \quad P_{q\gamma}(z) = \frac{1 + (1-z)^2}{z}\,.
\ee
Here,  $P_{q\gamma}(z)$ is the quark-to-photon splitting function with $z = k_\gamma^+/(k_\gamma^+ + q^+)$. The remaining kinematic variables are given as
\be
x_p = \frac{k_\gamma^+ + q^+}{P_p^+} \,, \qquad x_A= \frac{k_\gamma^- + q^-}{P_A^-}\,, \qquad Y_A = \log \frac{1}{x_A}\,,
\label{eq:kin}
\ee
with the projectile and target light-cone momenta in the center of mass frame given as $P_A^- = P_p^+ = \sqrt{\frac{s}{2}}$, where $s$ is the center-of-mass collision energy. Here, and in the following, the light-cone variables are defined as $p^{\pm} = (p^0 \pm p^3)/\sqrt{2}$ with the rapidity $\eta = \log(p^+/p^-)/2$. $\eta_\gamma$ and $\eta_h$ are the photon and the hadron rapidities, respectively. 

The function $\tilde{\calN}_{A,Y_A}(\kp)$ is the CGC dipole in the fundamental representation
\be
\tilde{\calN}_{A,Y_A}(\kp) = \int \rmd^2 \bperp \rme^{\rmi\kp\cdot \bperp} \tilde{\calN}_{A,Y_A}(\bperp)\,, \qquad  \tilde{\calN}_{A,Y_A}(\bperp) = \frac{1}{N_c} {\rm tr}_c \langle \tilde{U}(\bperp) \tilde{U}^\dag(0) \rangle_{Y_A}\,,
\label{eq:dip}
\ee
where $\tilde{U}(\bperp)$ is the fundamental light-like Wilson line arising from all order scattering on the dense target. The $k_\perp$-dependent gluon distribution $\varphi_{\rm DP}(Y,\kp)\sim \kp^2\tilde{\calN}_{Y}(\kp)$ in \eqref{eq:incgammahlo}, signifies a broadening of the collinear $2\to 2$ away side peak, that would be represented simply by a $\delta^{(2)}(\kp)$ (see \eqref{eq:pqcd}) and that is also shifted from $k_\perp = 0$ to $k_\perp \sim Q_S$. 

As mentioned in the introduction, a well known theoretical distinction of $\gamma h$ with respect to $hh$ correlations is that they probe fundamentally different gluon distributions \cite{Dominguez:2010xd,Dominguez:2011wm}. The leading order $\gamma h$ cross section is proportional to the gluon dipole distribution $\varphi_{\rm DP}(Y,\kp)$, while the leading order $hh$ cross section is proportional to a combination of the Weisz\" acker-Williams gluon distribution $\varphi_{\rm WW}(Y,\kp)$ and the dipole distribution \cite{Dominguez:2010xd,Dominguez:2011wm}. While these two distributions both display a high-$k_\perp$ perturbative $1/\kp^2$ tail, their behavior is completely different for low $k_\perp$ where we have 
\be
\varphi_{\rm DP}(Y,\kp)\sim \kp^2/Q_S^2\,, \qquad \varphi_{\rm WW}(Y,\kp)\sim \log(Q_S^2/\kp^2)\,.
\label{eq:dpww}
\ee
This prediction is particularly important at the away-side-peak, $\Delta\phi = \pi$, where small values of $k_\perp$ would be probed. In the case of $\gamma h$ correlations, $\varphi_{\rm DP}(Y,\kp)$ causes a dip in the cross section, see \cite{Rezaeian:2012wa} and also \cite{Rezaeian:2016szi,Goncalves:2020tvh,Stasto:2012ru,Basso:2015pba,Basso:2016ulb}, since the underlying partonic cross section in \eqref{eq:incgammahlo} is strictly vanishing as the kinematics condition $k_{\perp} = |\kgp + \qp|$ becomes satisfied. On the other hand, due to the low $k_\perp$ behavior of $\varphi_{\rm WW}(Y,\kp)$ the $hh$ production would rather level off to a constant at the away side peak.

In this paper, the $Y_A$ dependence of the gluon dipole follows the running coupling Balitsky-Kovchegov evolution equation \cite{Balitsky:1995ub,Kovchegov:1999yj,Balitsky:2006wa} (rcBK) which is a good approximation to the more general Jalilian-Marian-Iancu-McLerran-Weigert-Leonidov-Kovner (JIMWLK) evolution \cite{JalilianMarian:1997jx,JalilianMarian:1997dw,Iancu:2000hn,Iancu:2001ad} of the dipoles. As a general feature of small-$x$ evolution the distribution broadens as $x$ becomes smaller. The initial condition for the rcBK evolution is set at $x = x_0 = 0.01$ where the dipole is given by the anomalous dimension McLerran-Venugopalan (MV) model,  which we will call here the MV$^\gamma$ model. The initial condition for the rcBK evolution of the fundamental dipole is explicitly given by
\be
\tilde{\calN}_{Y_0}(\xp) = \exp\left\{- \frac{\left(x_\perp^2 Q_{S0}^2\right)^{\gamma}}{4} \log\left(\frac{1}{x_\perp \Lambda_{\rm IR}}+ \rme\right)\right\}\,,
\ee
where $Y_0 = \log 1/x_0$, $Q_{S0}$ is the initial saturation momentum, $\gamma$ is the anomalous dimension and $\Lambda_{\rm IR}$ is the IR cutoff of the model. We use the parameter set \cite{Albacete:2010sy} $\gamma = 1.119$, $(Q_{S,0}^p)^2 = 0.169 \, {\rm GeV}^2$, $\Lambda_{\rm IR} = 0.241 \, {\rm GeV}$. The rcBK equation provides the dipole distribution at $x < x_0$ while for $x > x_0$ we use the matching to the collinear gluon PDF as explained in \cite{Ma:2015sia}. The matching procedure fixes the proton radius to $R_p = 0.5257$ fm. For the nuclei we use $Q_{S0,A}^2 = c A^{1/3} Q^2_{S0,p}$ where $c\simeq 0.5$ \cite{Dusling:2009ni}. In our computations we set $Q_{S0,A}^2 = 3 Q^2_{S0,p}$ for heavy nuclei such as Pb and Au.

We will be computing the cross section for an isolated $\gamma h$ production, defined in a standard way through a fixed isolation cut in $\eta-\phi$ space around the photon, defined by $(\eta_\gamma,\phi_\gamma)$. An isolation cone, $R$, is introduced to cut out a region where $\sqrt{\Delta \eta^2 + \Delta\phi^2} < R$ in order to isolate the photon from any soft and collinear hadronic activity within $R$. Here $\Delta \eta = \eta_\gamma - \eta_h$ and $\Delta\phi = \phi_\gamma - \phi_h$. For the cross section at hand \eqref{eq:incgammahlo} this effectively suppresses its fragmentation component that would appear when $\gamma$ is being emitted collinearly to the underlying final state parton \cite{Jalilian-Marian:2012wwi}.

\subsection{Implementing the Sudakov effect}
\label{sec:sudakov}

The implementation of the Sudakov effect rests on the Collins-Soper-Sterman (CSS) or transverse momentum resummation formalism \cite{Collins:1984kg,Collins:2011zzd}. The inclusion of soft gluon radiation to all orders introduces a transverse momentum dependence into the distribution and fragmentation functions that follows the CSS evolution equation \cite{Collins:1984kg,Collins:2011zzd}. The result of CSS evolution is obtained in $\bperp$-space resulting in a compact formula for the so-called Sudakov factor. Refs.~\cite{Mueller:2012uf,Mueller:2013wwa} further demonstrated that Sudakov resummation can be accomplished on top of the small-$x$ resummation at the one-loop order. In the context of $\gamma h$ production considered here the relevant part of the cross section \eqref{eq:incgammahlo} gets modified through 
\be
\kp^2 \tilde{\calN}_{A,Y_A}(\kp) D_q(z_h,\mu^2) f_q(x_p,\mu^2) \to \int \rmd^2 \bperp \rme^{\rmi\kp\cdot \bperp} \partial^2_{\bperp}\tilde{\calN}_{A,Y_A}(\bperp)D_q(z_h,\mu_b^2) f_q(x_p,\mu_b^2)\, \rme^{-S_{\rm Sud}(\bperp,Q)}\,,
\label{eq:sud1}
\ee
where we multiplied by a prefactor $\kp^2$ to get the unintegrated gluon distribution\footnote{See Appendix \ref{sec:appa} for the corresponding collinear formula.}. In Eq.~\eqref{eq:sud1} the exponential $e^{-S_{\rm Sud}(\bperp,Q)}$ corresponds to the resummed contribution from soft-gluon radiations, %of multiple-gluon radiation branchings,  
where the Sudakov factor $S_{\rm Sud}(\bperp,Q)$,  has been introduced as the radiation kernel summed over $\mu_b$ up to the hard scale $Q$. The Sudakov factor is given by the following generic form \cite{Sun:2014gfa,Sun:2015doa}
\be
S_{\rm Sud}(\bperp,Q) = \int_{\mu_b^2}^{Q^2} \frac{d\bar{\mu}^2}{\bar{\mu}^2} \left[A \log\left(\frac{Q^2}{\bar{\mu}^2}\right) + B\right]\,,
\label{eq:sudpertcgc}
\ee
Here $\mu_b$ is a factorization scale according to the $b_*$-prescription \cite{Collins:1984kg}
\be
\mu_b  = \frac{2 \rme^{-\gamma_E}}{b_*} \,, \qquad b_*^2 = \frac{\bperp^2}{1 + \frac{\bperp^2}{b_{\rm max}^2}}\,,
\ee
where $b_{\rm max} = 1.5$ GeV$^{-1}$. With this prescription $\mu_b > 2 \rme^{-\gamma_E}/b_{\rm max}$ thus preventing the $\bperp$-integral from entering the non-perturbative region. For the hard scale we take $Q^2  = x_p x_A s$. The quantities $A$ and $B$ are channel dependent coefficients that can be computed in perturbation theory. For each initial and final state quark (or gluon) in the channel we have $A_q = \alpha_S C_F/(2\pi)$, $A_g = \alpha_S C_A/(2\pi)$, and
$B_q = -\alpha_S \frac{3}{2}C_F/(2\pi)$, $B_g = -\alpha_S 2C_A \beta_0/(2\pi)$, where $\beta_0 =\left(11 - \frac{2 N_f}{3}\right)/12$ at one-loop order \cite{Sun:2014gfa,Sun:2015doa}. 
According to Refs. \cite{Mueller:2012uf,Mueller:2013wwa},  the pertubative prefactor is absent in the single-log $B$ term for the case of any incoming small-$x$ gluon.  Consequently, we do not count the small-$x$ gluon in obtaining this coefficient. For the $qg \to q \gamma$ channel we should use
\be
A = 2A_q + A_g = \frac{\alpha_S(\bar{\mu}^2)}{\pi}\left( C_F + \frac{C_A}{2}\right)\,, \qquad B = 2 B_q = - \frac{\alpha_S(\bar{\mu}^2)}{\pi} \frac{3}{2} C_F\,.
\label{eq:absud}
\ee

In order to compensate for the missing effect at large $\bperp$ it is common to add a non-perturbative Sudakov factor $S_{\rm non-pert}(\bperp,Q)$ as
\be
S_{\rm Sud}(\bperp,Q) \to S_{\rm Sud}(\bperp,Q) + S_{\rm non-pert}(\bperp,Q)\,.
\ee
In this work we are using the parametrization \cite{Sun:2014dqm}
\be
S^q_{\rm non-pert}(\bperp,Q) = \frac{g_1}{2}\bperp^2 + \frac{1}{4}\frac{g_2}{2}\log\frac{Q^2}{Q^2_0}\log\frac{\bperp^2}{b_*^2}\,, \qquad S^g_{\rm non-pert}(\bperp,Q) = \frac{C_A}{C_F}S^q_{\rm non-pert}(\bperp,Q)\,,
\label{eq:sudnonpert}
\ee
where $g_1 = 0.212$ GeV$^2$, $g_2 = 0.84$, $Q_0^2 = 2.4$ GeV$^2$. Eq.~\eqref{eq:sudnonpert} should in principle be used for each quark and gluon in the initial or the final state.  The small-$x$ gluon already contains some non-perturbative information through the $k_\perp$-dependent distribution. %contained in the small-$x$ resummation of the CGC formalism,  which accounts for some non-perturbative effects. 
To avoid any possible double counting, the small-$x$ gluon is therefore dropped \cite{Zheng:2014vka,Stasto:2018rci} and we have
%Thus, in the $qg \to q\gamma$ channel we have 
$S_{\rm non-pert}(\bperp,Q) = 2 S^q_{\rm non-pert}(\bperp,Q)$.

Denoting
\be
W(z_h,x_p,\bperp,Q) \equiv \sum_q \frac{e_q^2 N_c}{8\pi^4} D_q(z_h,\mu_b^2)f_q(x_p,\mu_b^2)\rme^{-S_{\rm Sud}(\bperp,Q)-S_{\rm non-pert}(\bperp,Q)}\,,
\label{eq:W}
\ee
our final formula for the $\gamma h$ cross section reads
\be
\frac{\rmd \sigma}{\rmd^2 \kgp \rmd \eta_\gamma\rmd^2 \Php\rmd \eta_h} = (\pi R_A^2) \int_0^1 \frac{\rmd z_h}{z_h^2} \int\frac{\rmd^2 \kp'}{(2\pi)^2}W(z_h,x_p,\kp' - \kp,Q) \kp'^2\tilde{\calN}_{A,Y_A}(\kp')\hat{\sigma}\,,
\label{eq:incgammahlo2}
\ee
where we recall that $\kp = \kgp + \Php/z_h$.
The explicit computation of Eq.~\eqref{eq:incgammahlo2} is performed in momentum space through a convolution of the momentum space gluon distribution, $\kp^2 \tilde{\calN}_{A,Y_A}(\kp)$,  with the Fourier transform of $W(z_h,x_p,\bperp,Q)$. The latter is computed numerically with the algorithm in \cite{Kang:2019ctl}. We also remark that with resumming infinite number of soft gluons exact kinematics relations \eqref{eq:kin} are lost, and in practice \cite{Mueller:2016gko,Chen:2016vem} one resorts to an approximate relation $x_{p,A} = p_\perp (\rme^{\pm\eta_\gamma} + \rme^{\pm\eta_h})/\sqrt{s}$, with $p_\perp \equiv {\rm max}(k_{\gamma\perp},q_\perp)$.

According to \eqref{eq:W} the Sudakov factor brings an additional dependence of the imbalance $k_\perp$-distributions on the hard scale $Q$. As a typical result of CSS evolution, the $k_\perp$ distribution gets broadened as a function of the hard scale $Q$. This can be intuitively understood from the leading $\log^2(k_\perp^2/Q^2)$-dependence as the increase in the hard scale brings more opportunity for soft gluon radiation. In the CGC framework radiative processes are enhanced as $\log(1/x)$. Therefore, since $x \sim Q/\sqrt{s}$, an increase in the hard scale would lead to a narrower distribution. It is thus theoretically and phenomenologically interesting to check the interplay between the small-$x$ resummation and the Sudakov resummation.

\section{Numerical results and discussion}
\label{sec:results}

 %--- figure ---%
\begin{figure}
  \begin{center}
  \includegraphics[scale = 0.5]{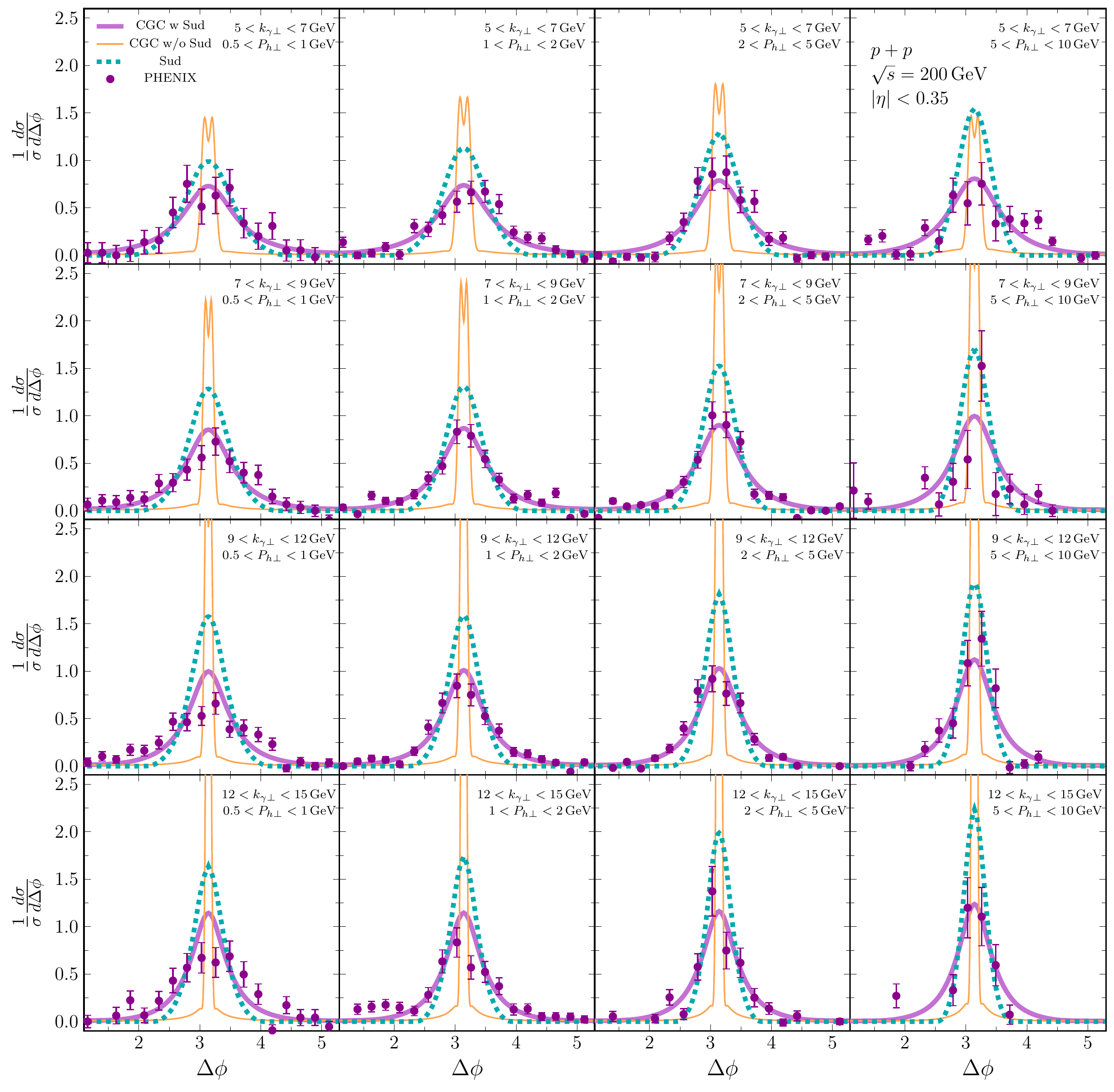}
  \end{center}
  \caption{Isolated $\gamma h^{\pm}$ angular distribution for mid-rapidity $pp$ collisions at $200$ GeV given in $k_{\gamma\perp}\times P_{h\perp}$ bins vs data from PHENIX \cite{PHENIX:2018trr}. The experimental data has been shifted for the underlying event. The solid magenta line corresponds to the combined CGC w Sud calculation, while the CGC w/o Sud and Sud computations correspond to the thin orange and dashed teal lines, respectively.}
  \label{fig:cpphenix200}
\end{figure}

In this Section we perform  
a numerical computation of the cross section for the isolated $\gamma h$ cross section based on \eqref{eq:incgammahlo2}. In what follows,  we abbreviate this result as as CGC w(ith) Sud(akov). To isolate the effects of both small-$x$ and Sudakov gluons and to gain further insight we will also show results from the CGC framework alone (Eq. \eqref{eq:incgammahlo}),  and from the cross-section in the collinear pQCD framework with only the Sudakov effect taken on account. For the latter we refer to \eqref{eq:colsud} in Appendix \ref{sec:appa}. We abbreviate these results as CGC w/o Sud and Sud, respectively. 

To keep the discussion within phenomenological reach, we will mostly study systems and kinematic windows already probed by different experimental collaborations.  Firstly,  the PHENIX experiment measured isolated $\gamma h^\pm$ production in $pp$ collisions at $\sqrt{s} = 200$ GeV \cite{PHENIX:2018trr}, covering the kinematic range
\be
|\eta_\gamma| < 0.35 \,, \qquad 5 \, {\rm GeV} < k_{\gamma\perp} < 15 \, {\rm GeV} \,, \qquad |\eta_h| < 0.35 \,, \qquad 0.5 \, {\rm GeV} < P_{h\perp} < 10 \, {\rm GeV}\,,
\label{eq:phenix200}
\ee
where an isolation cut $R = 0.3$ has been applied to the photon. For $\sqrt{s} = 510$ GeV collisions \cite{PHENIX:2016zxb} we have the following kinematics
\be
|\eta_\gamma| < 0.35 \,, \qquad 7 \, {\rm GeV} < k_{\gamma\perp} < 15 \, {\rm GeV} \,, \qquad |\eta_h| < 0.35 \,, \qquad 0.7 \, {\rm GeV} < P_{h\perp} < 10 \, {\rm GeV}\,,
\label{eq:phenix510}
\ee
and $R = 0.4$. The kinematics \eqref{eq:phenix200} and \eqref{eq:phenix510} has been further separated in $k_{\gamma\perp} \times P_{h\perp}$ bins as indicated in Figs.~\ref{fig:cpphenix200} and \ref{fig:cpphenix510}, respectively.

The ALICE experiment measured isolated $\gamma h^{\pm}$ production in $pp$ and $p$Pb collisions at $\sqrt{s} = 5.02$ TeV \cite{Acharya:2020sxs} with the following kinematics
\be
|\eta_\gamma| < 0.67 \,, \qquad 12 \, {\rm GeV} < k_{\gamma\perp} < 40 \, {\rm GeV} \,, \qquad |\eta_h| < 0.8 \,, \qquad 0.5 \, {\rm GeV} < P_{h\perp} < 10 \, {\rm GeV}\,,
\ee
and $R = 0.4$. The transverse hadron momenta $P_{h\perp}$ is further distributed into bins as indicated in Fig.~\ref{fig:cpalice}.

Before considering the results we make a general remark about our computation based on \eqref{eq:incgammahlo2}. In the kinematics region where the transverse momentum of the final state $\kgp + \Php$ is large, and where also $(\kgp + \Php)^2 \ll Q^2$, the perturbative Sudakov factor, \eqref{eq:sudpertcgc}, would dominate the overall two-particle momentum imbalance in the cross section. While the non-perturbative Sudakov factor is necessary to carry out the $\bperp$-integral, it is irrelevant for the $k_\perp$-spectrum, as demonstrated in \cite{Mueller:2016gko}. However, in our computations this is not completely the case, since for RHIC kinematics we have $Q \sim 7-21$ GeV, where the non-perturbative Sudakov factor should also play a role. At the LHC $Q \sim 17-55$ GeV, but also $k_\perp$ is possibly larger due to the more asymmetric $\gamma h$ momenta configuration.

We will first show our results for the angular distributions and compare them with the data from RHIC and the LHC. In our compuations we focus only on the shape of the $\gamma h$ yield and thus normalize both our theoretical curves and the experimental data to unity. Similar procedure has been employed e.~g. in \cite{Chen:2016vem,Chen:2018fqu,Stasto:2018rci}. We compute as well the normalized out-of-plane transverse momentum distributions and extract their Gaussian widths and compare both with the data from RHIC. We also provide the predicted nuclear modifications by showing the CGC w Sud computation at RHIC and the LHC.

\subsection{Angular distributions}

 %--- figure ---%
\begin{figure}
  \begin{center}
  \includegraphics[scale = 0.4]{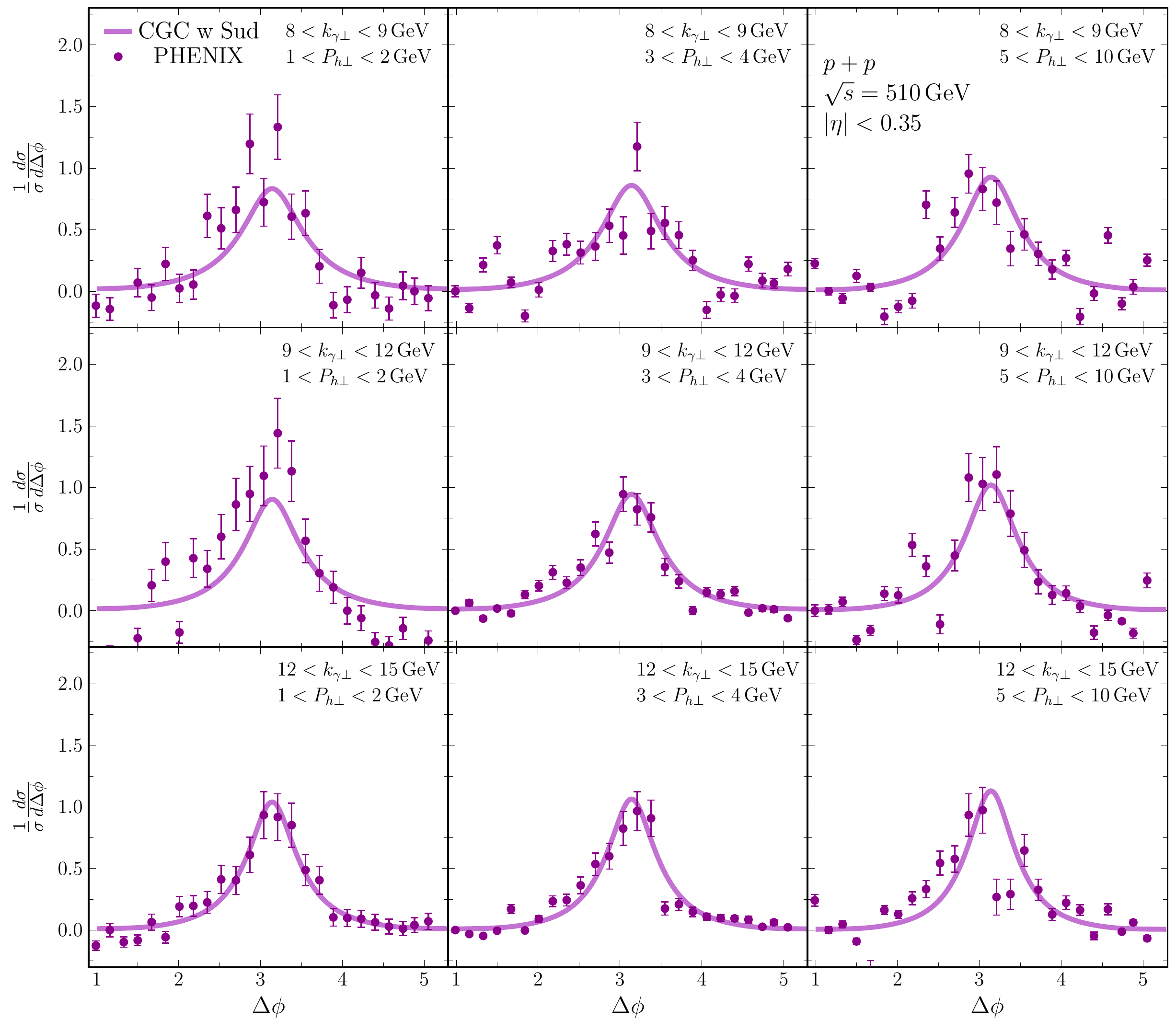}
  \end{center}
  \caption{Isolated $\gamma h^{\pm}$ angular distribution for mid-rapidity $pp$ collisions at $510$ GeV given in $k_{\gamma\perp}\times P_{h\perp}$ bins vs data from PHENIX \cite{PHENIX:2016zxb}. The experimental data has been shifted for the underlying event.}
  \label{fig:cpphenix510}
\end{figure}

In Fig.~\ref{fig:cpphenix200} we compare our CGC w Sud result with the $200$ GeV $pp$ data at RHIC \cite{PHENIX:2018trr} and find a fair agreement in most of the bins. Here the experimental data have been shifted vertically for the underlying event, that is, by the background of uncorrelated $\gamma h^\pm$ pairs \cite{PHENIX:2018trr}. The CGC w Sud results are also compared with a CGC w/o Sud computation. The away side peak from the CGC w/o Sud results is clearly too narrow to be able to describe the data. As argued previously, the dip at $\Delta \phi = \pi$, and the resulting double peak structure around it, for the CGC w/o Sud computation is due to the low $k_\perp$ behavior of the dipole gluon distribution $\varphi_{\rm DP}(Y,\kp) \sim \kp^2 /Q_S^2$. For this kinematics the double peak is strongly focused in a narrow region around $\Delta \phi = \pi$. Let us stress again here that the presence of a double peak in general is a robust prediction \cite{Rezaeian:2012wa,Rezaeian:2016szi,Goncalves:2020tvh,Stasto:2012ru,Basso:2015pba,Basso:2016ulb} of the leading order $\gamma h$ production in CGC. From the PHENIX data alone it is difficult to find support for this feature, though it might be simply missed by the experimental resolution. In any case, our prediction is that, for the kinematics considered here, the double peak is completely washed away by including the Sudakov effect. It is instructive to also compare to a Sud only result based on the leading order $qg \to q\gamma$ collinear formula. While the Sud result gets closer to the data than CGC w/o Sud in general, best results are obtained when both the CGC and the Sudakov effects are taken into account.

 %--- figure ---%
\begin{figure}
  \begin{center}
  \includegraphics[scale = 0.5]{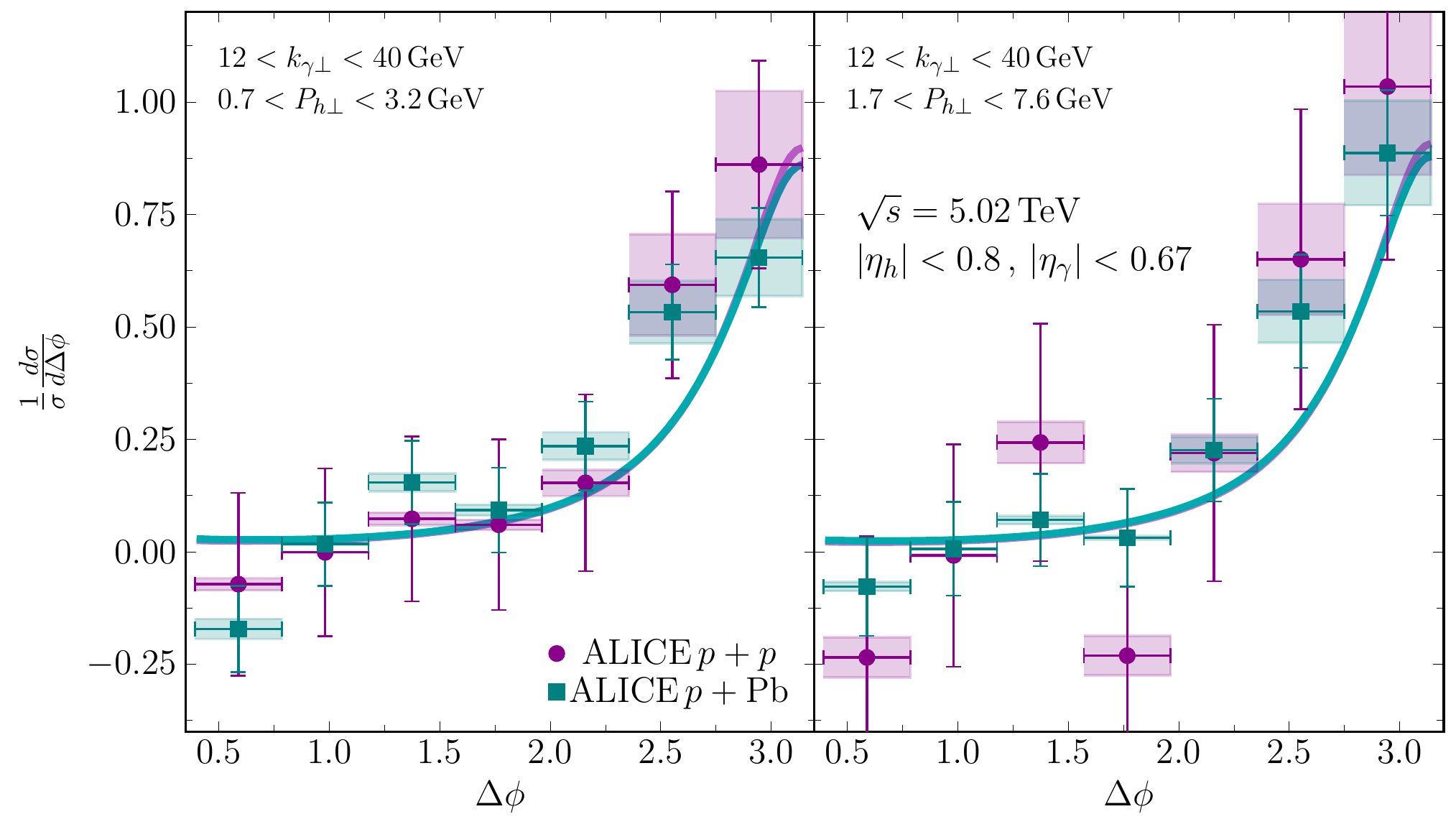}
  \end{center}
  \caption{Isolated $\gamma h^{\pm}$ angular distribution for mid-rapidity $pp$ and $p$Pb collisions at $5.02$ TeV for $12 \, {\rm GeV} < k_{\gamma\perp} < 40 \, {\rm GeV} $ for three bins in $P_{h\perp}$ vs data from ALICE \cite{Acharya:2020sxs}. The magenta points (lines) correspond to $pp$ collisions while the teal squares (lines) correspond to a $p$Pb collision.}
  \label{fig:cpalice}
\end{figure}

In Fig.~\ref{fig:cpphenix510} we compare the CGC w Sud computation with the $510$ GeV $pp$ data from RHIC \cite{PHENIX:2016zxb} where again we find overall good agreement with the data. In Fig.~\ref{fig:cpalice} we compare with the $5.02$ TeV $pp$ and $p$Pb data from LHC \cite{Acharya:2020sxs} and find good agreement with both the $pp$ and the $p$Pb data. Due to asymmetric kinematics in Fig.~\ref{fig:cpalice} our results show only a small nuclear effect. Below (see Fig.~\ref{fig:alicepred}) we look at more symmetric configurations.

 %--- figure ---%
\begin{figure}
  \begin{center}
  \includegraphics[scale = 0.6]{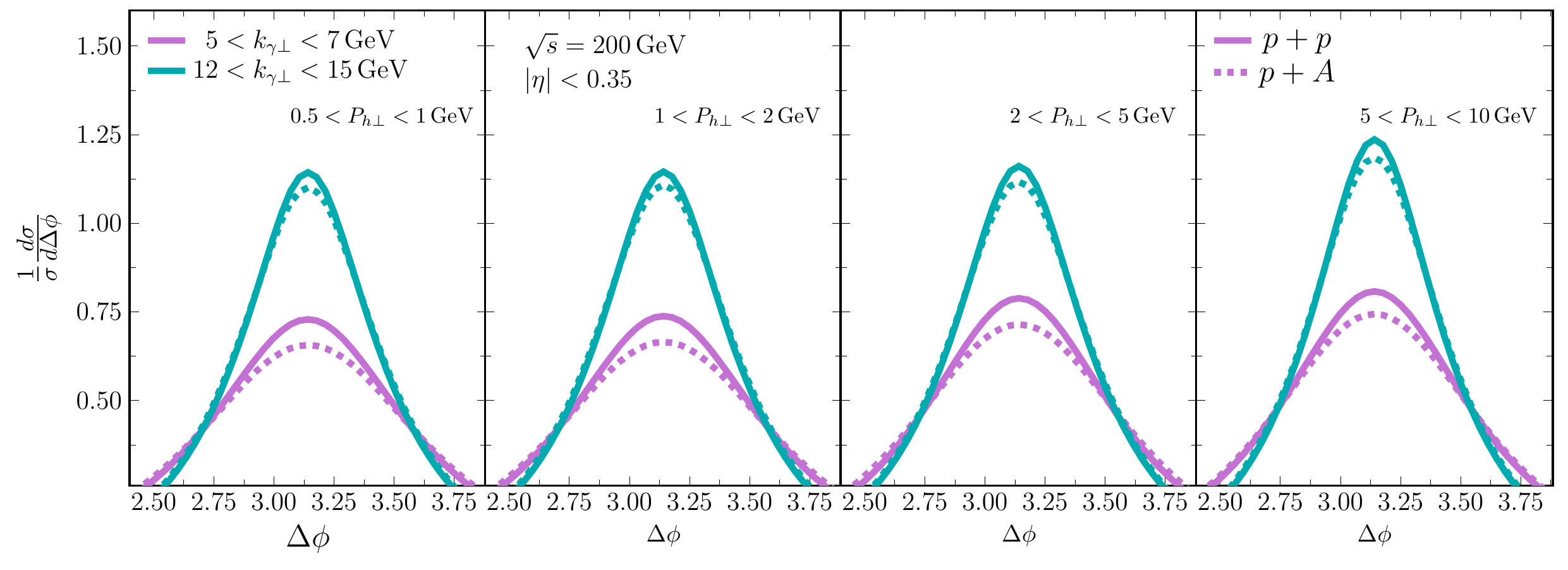}
  \end{center}
  \caption{Isolated $\gamma h^{\pm}$ angular distribution for mid-rapidity $pp$ and $pA$ collisions at $200$ GeV. The higher curve (and narrower peak) correspond to the  $5<k_{\gamma\perp}< 7$ GeV bin, while the lower (wider) peak corresponds to a harder photon momentum bin, $12<k_{\gamma\perp}< 15$ GeV.}
  \label{fig:phenix200pred}
\end{figure}

In Fig.~\ref{fig:phenix200pred} we make predictions for $pA$ collisions at $200$ GeV by considering a subset of bins from Fig.~\ref{fig:cpphenix200}. We pick up the smallest and the largest $k_{\gamma\perp}$ bin from Fig.~\ref{fig:cpphenix200} and distribute the results along $P_{h\perp}$ bins. For comparison, we also repeat the $pp$ results from Fig.~\ref{fig:cpphenix200}. It is useful at this point to discuss the systematics across bins. Lets compare two $pp$ (or $pA$) curves corresponding to the smallest and the largest $k_{\gamma\perp}$, for a fixed $P_{h\perp}$. We see that as $k_{\gamma\perp}$ (the transverse momentum of the trigger particle) is increased, the away side peak gets narrower. We can understand this in a intuitive way as follows. At large trigger $k_{\gamma\perp}$ also the momentum imbalance $k_\perp$ eventually increases ($P_{h\perp}$ is held fixed) and so the Sudakov logs become less prominent. Note that for high $k_{\gamma\perp}$-kinematics the details of the non-perturbative Sudakov factor should be negligible. At the same time, nonlinear effects from the CGC also play less of a role as with high trigger $k_{\gamma\perp}$ we are probing the perturbative tail of the gluon transverse momentum distribution. Therefore, we expect the probability for a  high-$k_{\gamma\perp}$ trigger photon to scatter at an angle $\Delta\phi \neq \pi$ to be strongly suppressed, explaining the narrower shape.  Likewise,  for smaller $k_{\gamma\perp}$ the away side peak will get broadened. Our second point concerns the nuclear effect that is visible by a comparison of the full ($pp$) and the dashed ($pA$) curves in Fig.~\ref{fig:phenix200pred}. Due to a larger saturation scale in the nuclei than in the proton target we observe a suppression of the away side peak in $pA$ in comparison to $pp$ for all kinematic bins considered.

 %--- figure ---%
\begin{figure}
  \begin{center}
  \includegraphics[scale = 0.4]{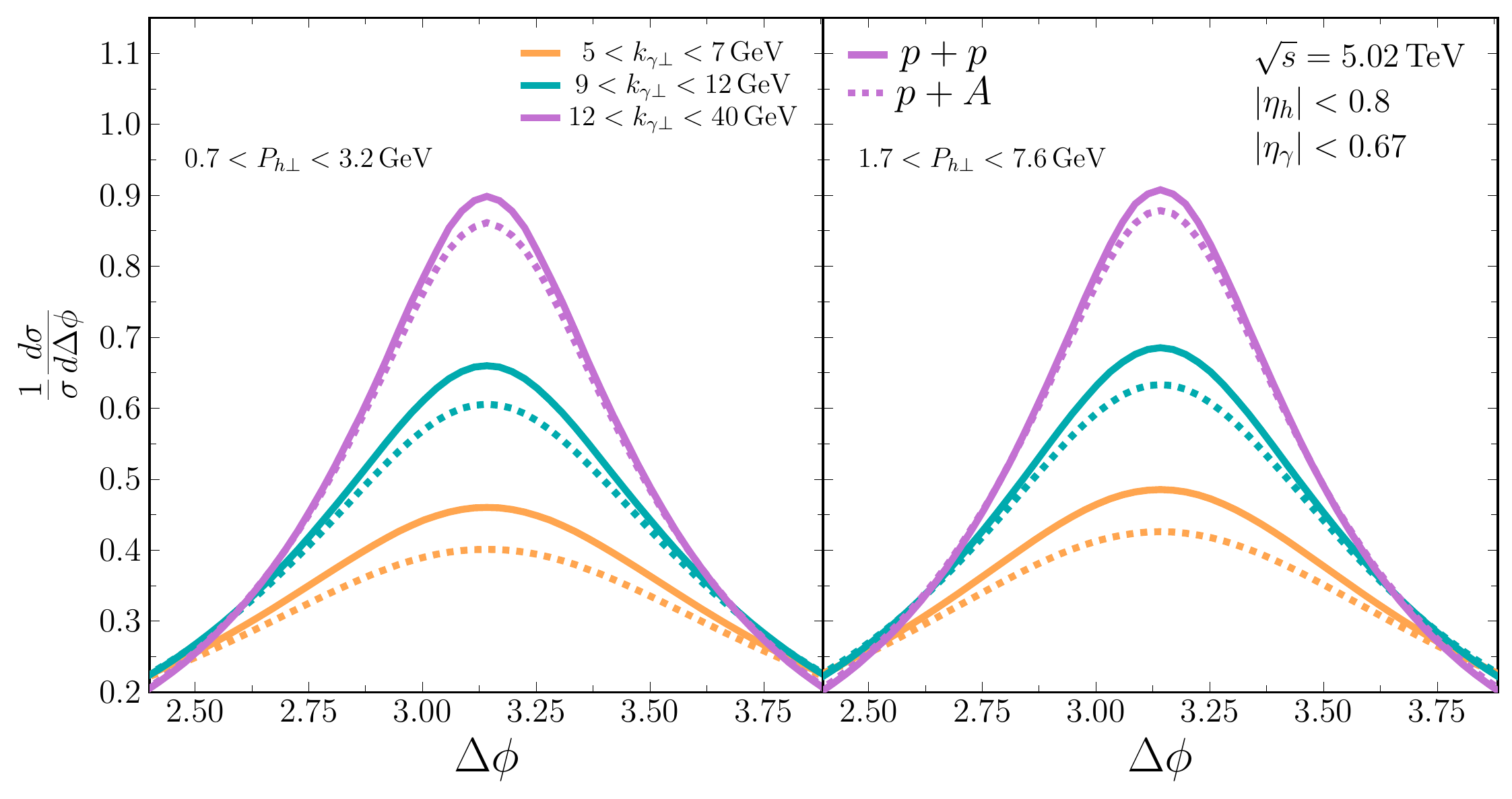}
  \end{center}
  \caption{Isolated $\gamma h^{\pm}$ angular distribution for mid-rapidity $pp$ (solid) and $pA$ (dashed) collisions at $5.02$ TeV. Each $pp$ and $pA$ curve pair corresponds to a different binning of photon momentum. In descending order, the pair correspond to bins of $5<k_{\gamma\perp}< 7$ GeV,  $9<k_{\gamma\perp}< 12$ GeV ,  and $12<k_{\gamma\perp}< 15$ GeV.}
    \label{fig:alicepred}
\end{figure}

In Fig.~\ref{fig:alicepred} we present projections for mid-rapidity $pp$ and $pA$ collisions at the LHC energy of $5.02$ TeV. We keep the associated hadron bins from Fig.~\ref{fig:cpalice} and lower the trigger photon $k_{\gamma\perp}$ in order to reach a more symmetric configuration. The imbalance momentum of the $\gamma h^{\pm}$ system thus approaches the saturation scale at the away side peak leading to a more substantial nuclear effect as can be seen from Fig.~\ref{fig:alicepred}.

\subsection{Out-of-plane transverse momentum distributions}

In addition to the angular distributions, PHENIX also measured \cite{PHENIX:2018trr,PHENIX:2016zxb} the so-called out-of-plane $p_{\rm out}$ distributions where $p_{\rm out} \equiv P_{h\perp}\sin(\Delta\phi)$, binned as a function of $x_E \equiv -\kgp\cdot \Php/\kgp^2 = -P_{h\perp}\cos(\Delta\phi)/k_{\gamma\perp}$. A quick computation shows that $p_{\rm out}^2 = z_h^2 \kp^2 - \kgp^2 (1-x_E/z_h)^2$ at the leading order. Close to the away side peak $\Delta \phi =  \pi$ we have $x_E \simeq z_h$ and so $p_{\rm out} \simeq z_h k_\perp$.  By binning the result in $x_E$, we can appreciate that the $p_{\rm out}$-distributions serve as a proxy for the intrinsic $k_\perp$ distributions \cite{PHENIX:2018trr,Osborn:2018bwn}.

 %--- figure ---%
\begin{figure}
  \begin{center}
  \includegraphics[scale = 0.6]{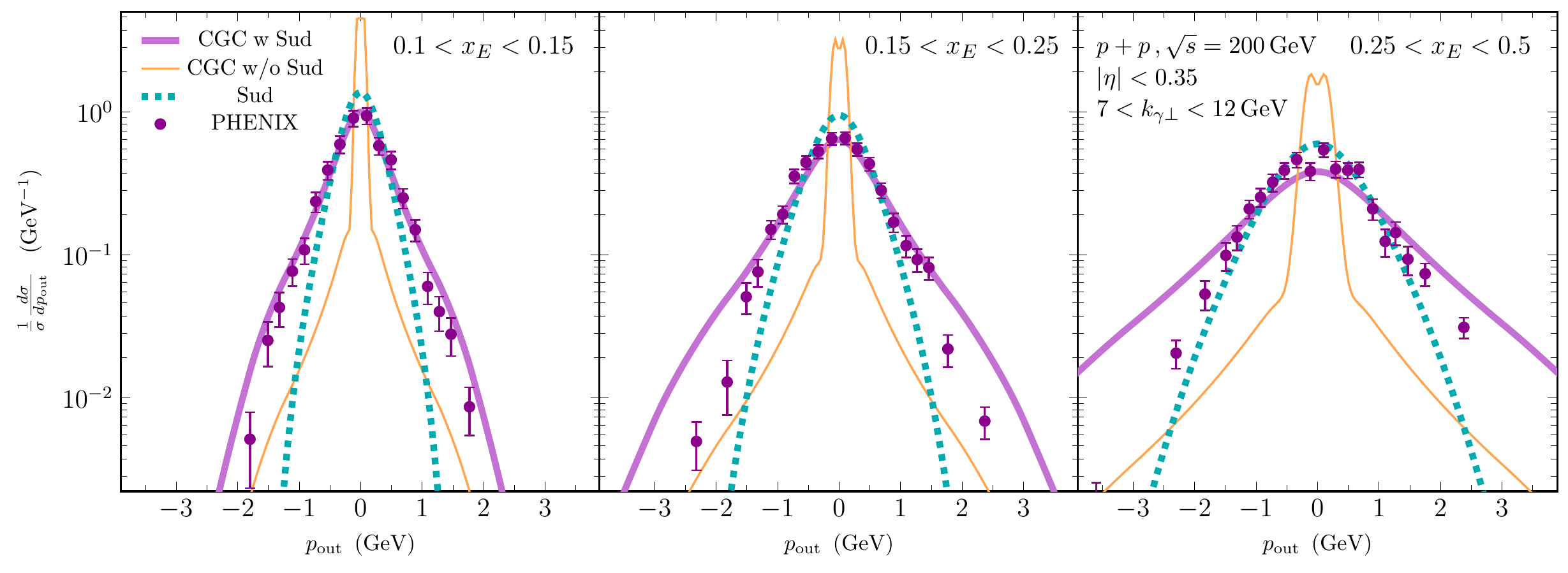}
  \end{center}
  \caption{$p_{\rm out}$ distributions for mid-rapidity $pp$ collisions at $\sqrt{s} = 200$ GeV compared to the data from PHENIX \cite{PHENIX:2018trr} in three $x_E$ bins, $0.1<x_E<0.15$ (left), $0.15<x_E<0.25$ (center), and $0.25<x_E<0.5$ (right).The solid magenta line corresponds to the combined CGC w Sud calculation, while the CGC w/o Sud and Sud computations correspond to the thin orange and dashed teal lines, respectively. }
  \label{fig:poutdist}
\end{figure}

In Fig.~\ref{fig:poutdist} we compare our results with the PHENIX 200 GeV data \cite{PHENIX:2018trr} in $x_E$ bins. The CGC w Sud results show a good agreement with the data in the small to moderate $p_{\rm out}$ region (up to $\sim 1-2$ GeV), and also in the large $p_{\rm out}$ region for the $x_E$ bin $0.1 < x_E < 0.15$. For the remaining two bins, the small to moderate $p_{\rm out}$ region ($p_{\rm out} \sim 1-2$ GeV) is also nicely described by our result, while as $p_{\rm out}$ increases our results tend to overshoot the data. For comparison, the CGC w/o Sud computation is also shown where similar conclusions hold as for the angular distributions: the CGC w/o Sud computation produces a too narrow distribution which cannot be accommodated within the experimental data. In addition, the CGC w/o Sud computation predicts a double peak in the distribution concentrated in a narrow region around $p_{\rm out} = 0$. Again, the present PHENIX data sets alone do not allow us to judge clearly whether this feature is supported or not.
We can only underline the importance of a more complete computation, which includes the Sudakov resummation, where a double peak structure is not present for these kinematics. For completeness we also show the results of a Sud computation which seems to do a good job in the moderate $p_{\rm out}$ region, but tends to undershoot the data for large $p_{\rm out}$.

 %--- figure ---%
\begin{figure}
  \begin{center}
  \includegraphics[scale = 0.4]{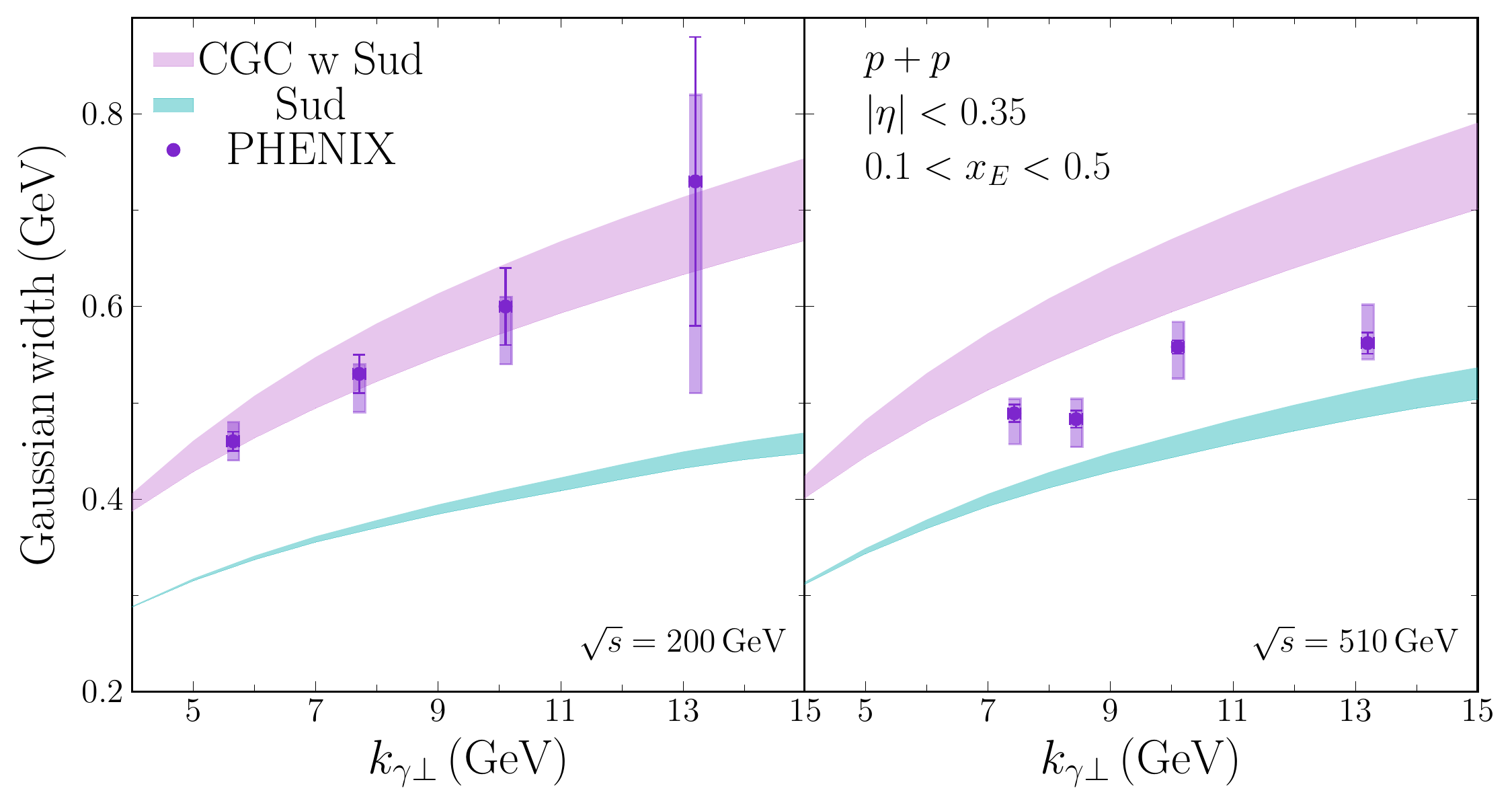}
  \end{center}
  \caption{Gaussian widths of the $p_{\rm out}$ distributions as a function of the hard scale $k_{\gamma\perp}$ for mid-rapidity $pp$ collisions. The $pp$ results at $\sqrt{s} = 200$ GeV and $\sqrt{s} = 510$ GeV are compared to the data from PHENIX \cite{PHENIX:2018trr}. The magenta bands correspond to the CGC w Sud computation, while the teal bands correspond to the Sud computation.}
  \label{fig:ktbroad}
\end{figure}

To have a closer look at the small $p_{\rm out}$ region, PHENIX extracted the Gaussian widths of the $p_{\rm out}$-distributions across $0.1 < x_E < 0.5$ from Fig.~\ref{fig:poutdist} by assuming a Gaussian-like shape in the range $-1.1 \, {\rm GeV}<p_{\rm out} < 1.1 \, {\rm GeV}$. The systematic error is estimated by varying this range by $\pm 0.2$ GeV \cite{PHENIX:2018trr}.
Using the same procedure we compute the Gaussian widths of the $p_{\rm out}$ distributions from Fig.~\ref{fig:poutdist}. Our results are shown in Fig.~\ref{fig:ktbroad} as a function of the hard scale $k_{\gamma\perp}$ in comparison to the PHENIX 200 GeV and 510 GeV $pp$ data. The theoretical bands correspond to varying the $p_{\rm out}$ Gaussian fit by $\pm 0.2$ GeV as in \cite{PHENIX:2018trr}. We see that the best description of both the 200 GeV and the 510 GeV data is obtained with the CGC w Sud computation. In Fig.~\ref{fig:ktbroad} we also plot Sud results for the Gaussian widths and find that they are below the PHENIX data. The CGC w/o Sud widths are not shown - due to the double peak structure the behavior is clearly not Gaussian-like for small $p_{\rm out}$. But even if we choose to ignore this issue, based on the fact that the double peak is rather narrow for this particular kinematics, it is visible already by the naked eye from Fig.~\ref{fig:poutdist} that the distribution are too narrow in comparison to the data.

 %--- figure ---%
\begin{figure}
  \begin{center}
  \includegraphics[scale = 0.6]{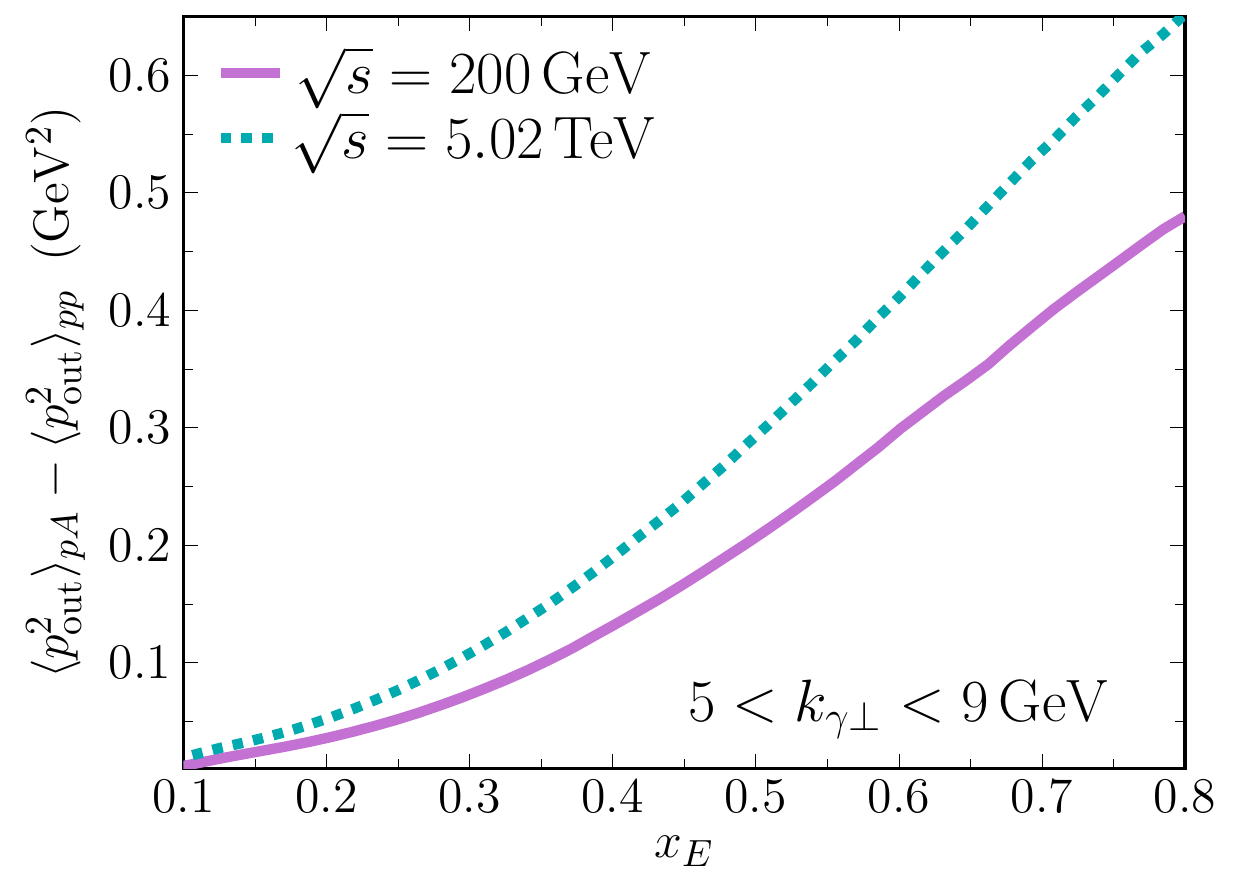}
  \end{center}
  \caption{The difference between $pA$ and $pp$ Gaussian widths squared of the $p_{\rm out}$ distributions as a function of $x_E$ at $\sqrt{s}=200$ GeV and $\sqrt{s}=5.02$ TeV collision energies.}
  \label{fig:ktbroad2}
\end{figure}

In Fig.~\ref{fig:ktbroad2}, as a measure of the nuclear effect, we show our predictions for the difference in the Gaussian widths squared computed in mid-rapidity $pA$ and in $pp$ at RHIC (left) and at the LHC (right) kinematics. For convenience we denote this quantity as $\langle p_{\rm out}^2  \rangle_{pA} - \langle p_{\rm out}^2 \rangle_{pp}$, though the meaning of the ``averaging" procedure $\langle \,\, \rangle$ is in the sense of fitting the small $p_{\rm out}$ region with a Gaussian - same as we used for Fig.~\ref{fig:ktbroad}. Our choice of kinematics matches the one used in the preliminary RHIC $p$Pb vs. $pp$ results in \cite{Osborn:2017ndr,Osborn:2018bwn,Aidala:2019sok} with the hard scale $5 \, {\rm GeV}<k_{\gamma\perp} < 9 \, {\rm GeV}$ and we plot our results as a function of $x_E$. We are tempted to compare the $200$ GeV results from Fig.~\ref{fig:ktbroad2} with the preliminary result in Refs.~\cite{Osborn:2017ndr,Osborn:2018bwn,Aidala:2019sok} that shows up to about $0.1$ GeV$^2$ broader widths in $pA$ that in $pp$, for $x_E < 0.4$, albeit for $\pi^0 h^\pm$ production. In our $\gamma h^\pm$ computation we find $\lesssim 0.15$ GeV$^2$ for $x_E < 0.4$.  Increasing the energy to $5.02$ TeV we find that the difference in the $pA$ vs. $pp$ Gaussian widths grows, becoming more pronounced at large $x_E$.

\section{Conclusions}
\label{sec:conc}

In this paper we have numerically computed the isolated photon-hadron production cross section based on the leading order CGC formula with the Sudakov resummation explicitly taken into account for the first time. We have demonstrated that both the CGC and the Sudakov effects are important to obtain a reasonable description of the data at RHIC and the LHC. We have provided predictions for the nuclear suppression around the away side peak $\Delta \phi = \pi$ and the nuclear broadening of the intrinsic transverse momentum distributions at RHIC and at the LHC. While the PHENIX collaboration published only $pp$ data, the ALICE collaboration has results also for $pp$ and $p$Pb. Unfortunately, there is no clear evidence of a nuclear effect from the data alone. This might be due to asymmetric $k_{\gamma\perp} \times P_{h\perp}$ binning by ALICE and we argue that in more symmetric configurations the nuclear effect would be more apparent.
In addition to $\gamma h^\pm $ production the PHENIX experiment measured also $\pi^0 h^\pm$ production \cite{PHENIX:2016zxb,PHENIX:2018trr}. The data seems to indicate a suppression in the $\gamma h^\pm$ vs. $\pi^0 h^\pm$ production in the bins that are close to the away side peak. It would be interesting to make a detailed side-by-side comparison of $\gamma h^\pm$ vs. $\pi^0 h^\pm$ production in CGC.

As a future work we plan to take into account next-to-leading order corrections which bring $2 \to 3$ partonic processes into play. Firstly, already in the collinear framework, the presence of an additional (unobserved) parton present in the final state naturally disrupts the $\gamma h$ back-to-back kinematics. As the Sudakov effect is associated with soft gluon branching it is most important in the region close to the away side peak with the momentum imbalance $k_\perp = |\qp + \kgp|$ such that $k_\perp \ll Q$. By contrast, the $2 \to 3$ processes are genuine hard branchings, that are able to support large momentum imbalances $k_\perp \sim Q$. Therefore, next-to-leading order corrections and Sudakov effects are important to get a more complete phase-space picture of $\gamma h$ correlations. This has recently been applied in describing hard-$p_\perp$ $\gamma$-jet data in $pp$, see e.~g.~\cite{Jezo:2016ypn,Klasen:2017dsy} and also \cite{Chen:2018fqu}. We also mention here recent works on transverse momentum resummation in $\gamma$-jet production that also takes into account a preferred direction set by the final state jet \cite{Buffing:2018ggv,Hatta:2021jcd}.

In the small-$x$ kinematics region with $k_\perp \sim Q_S$ we are interested, a complete next-to-leading order treatment would include considering the $g g \to q\bar{q}\gamma$ \cite{Benic:2016uku,Benic:2016yqt} and $qg \to qg \gamma$ channels \cite{Altinoluk:2018uax,Altinoluk:2018byz} (see also \cite{Roy:2018jxq,Roy:2019hwr} for related higher order computations in $eA$ collisions). With the current photon detectors in the mid-rapidity region for both RHIC and LHC, as a first step, the inclusion of the $g g \to q\bar{q}\gamma$ channel might be enough. Considering planned forward upgrades \cite{ALICE:2020mso,Berti:2021azk}, where the isolated photon signal would be more favourably extracted from the $\pi^0$ background, taking into account also the $qg \to qg \gamma$ channel becomes important.

\acknowledgments

We thank Abhiram Kaushik for discussions. S.~B. thanks Yoshitaka Hatta, Joseph Osborn, Sebastian Sapeta and Shu-yi Wei for very useful correspondences. S.~B. and A.~P. are supported by the Croatian Science Foundation (HRZZ) no. 5332 (UIP-2019-04).  This project was supported by  the Deutsche Forschungsgemeinschaft (DFG, German Research Foundation) – Project number 315477589 – TRR 211.  

\appendix

\section{Collinear formula}
\label{sec:appa}

Here we recall the collinear formula for the $pt \to h\gamma X$ cross section. We have
\be
\frac{\rmd \sigma}{\rmd^2 \kgp \rmd \eta_\gamma\rmd^2 \Php\rmd \eta_h} = \sum_q e_q^2 \int \frac{\rmd z_h}{z_h^2} D_q(z_h,\mu^2) x_p f_q(x_p,\mu^2)x_A f_g(x_A,\mu^2)\alpha_S\hat{\sigma}\delta^{(2)}(\kp)\,,
\label{eq:pqcd}
\ee
where for the $qg\to q\gamma$ channel \cite{Ellis:1996mzs} we have\footnote{The $q\bar{q} \to g \gamma$ channel is negligible in this kinematics region, see for example Fig.~12 in \cite{PHENIX:2016zxb}.}
\be
\hat{\sigma} = \frac{\alpha_e}{N_c \hat{s}^2}\left(-\frac{\hat{u}}{\hat{s}} - \frac{\hat{s}}{\hat{u}}\right)\,,
\label{eq:pqcd2}
\ee
with the conventional Mandelstam variables $\hat{s} = (p + k)^2$, $\hat{t} = (k - k_\gamma)^2$ and $\hat{u} = (p - k_\gamma)^2$. The parton momenta fractions are fixed as $x_p = (k_\gamma^+ + q^)/P_p^+$ and $x_A = (k_\gamma^- + q^-)/P_A^-$. Taking into account the Sudakov resummation yields the following result
\be
\begin{split}
\frac{\rmd \sigma}{\rmd^2 \kgp \rmd \eta_\gamma\rmd^2 \Php\rmd \eta_h} = \sum_q e_q^2 \int \frac{\rmd z_h}{z_h^2} \int \frac{\rmd^2 \bperp}{(2\pi)^2}& \rme^{\rmi \kp \cdot \bperp} D_q(z_h,\mu_b^2)  x_p f_q(x_p,\mu_b^2)x_A f_g(x_A,\mu_b^2)\\
&\times \alpha_S(\mu_b^2)\hat{\sigma} \rme^{-S_{\rm Sud}(\bperp,Q) - S_{\rm non-pert}(\bperp,Q)}\,.
\label{eq:colsud}
\end{split}
\ee
The Sudakov factor $S_{\rm Sud}(\bperp,Q) + S_{\rm non-pert}(\bperp,Q)$ contains perturbative and non-perturbative pieces as in \eqref{eq:W}. The perturbative piece has the same form as in \eqref{eq:sudpertcgc}, however,  because of the initial stage gluon,  in the collinear limit one has $B = 2 B_q + B_g$ (the double-log $A$-coefficient remains the same as in \eqref{eq:absud}). In the non-perturbative piece we have now $S_{\rm non-pert}(\bperp,Q) = 2S^q_{\rm non-pert}(\bperp,Q) + S^g_{\rm non-pert}(\bperp,Q)$.

It is useful to explicitly confirm that we can recover \eqref{eq:pqcd} from \eqref{eq:incgammahlo} in the leading twist limit. First lets rewrite \eqref{eq:pqcd2} in a more convenient form. Using $\hat{u} = - 2p\cdot k_\gamma = -\kgp^2/z $ and $\hat{s} = 2 q\cdot k_\gamma = \kgp^2/z(1-z)$ we get
\be
\hat{\sigma} = \frac{\alpha_e}{2 N_c}\frac{P_{q\gamma}(z)}{q\cdot k_\gamma}\frac{z^2}{\kgp^2}\,.
\ee
Recall now that the transverse momentum dependent gluon distribution $\varphi(Y,\kp)$ function is related to the adjoint dipole $\calN_Y(\kp)$ as \cite{Blaizot:2004wu}
\be
\varphi(Y,\kp) = (\pi R_A^2)\frac{N_c \kp^2}{4 \alpha_S} \calN_{Y}(\kp)\,.
\ee
Integrating $\varphi(Y,\kp)$ over $\kp$ returns the gluon distribution
\be
x f_g(x) = \frac{1}{\pi^2}\int \frac{\rmd^2 \kp}{(2\pi)^2}  \varphi(Y,\kp)\,,
\ee
which may be formally inverted as $\varphi(Y,\kp) = x f_g(x) \pi^2 (2\pi)^2 \delta^{(2)}(\kp)$. In the large $N_c$ limit, the adjoint dipole and the fundamental dipole in coordinate space are related as $\calN_Y(\bperp) = \tilde{\calN}_Y^{C_F/C_A}(\bperp)$, see e.~g. \cite{Kovchegov:2001sc}. In the leading twist approximation this effectively becomes $\calN_Y(\bperp) \simeq 2 \tilde{\calN}_Y(\bperp)$ and so we can write (in momentum space)
\be
\tilde{\calN}_Y(\kp) \simeq x f_g(x)\frac{8\pi^4 \alpha_S}{N_c}\delta^{(2)}(\kp)\,.
\ee
Then, Eq.~\eqref{eq:incgammahlo} yields Eq.~\eqref{eq:pqcd}.

\bibliographystyle{utphys}
\bibliography{references}

%\section*{References}
%\bibliographystyle{elsarticle-num}
%\bibliography{references}{}

\end{document}